\documentstyle[12pt]{article}
\newcommand{\cn}{\mathop{\rm cn}\nolimits}
\newcommand{\real}{\hbox{\rm I\hskip-2pt\bf R}}
\newcommand{\natur}{\hbox{\cal I\hskip-2pt\bf N}}
\newcommand{\natu}{\hbox{\normalsize\cal I\hskip-2pt\bf N}}

\begin{document}
\thispagestyle{empty}

\begin{center}
\Large \bf
Construction of Doubly Periodic Solutions via  the
               Poincar\'e--Lindstedt Method in the case of Massless\\
               $\varphi^4$ Theory\\[8mm]

\large\bf  Oleg~A.~Khrustalev\\[3mm]
{\it Bogoliubov \ Institute for Theoretical Problems of Microphysics,\\
         Moscow State University;\\
         Department of Physics,\\
	 Moscow State University,\\[1mm]
         Vorobievy Gory, Moscow, 119899, RUSSIA}\\[2mm]
        {\normalsize \bf E-mail: khrust@sunny.bog.msu.su\\[3mm]
           \rm and}\\[4mm]
 Sergey~Yu.~Vernov\\[5mm]
         {\it  Skobeltsyn \ Institute for Nuclear Physics,\\
         Moscow State University,\\[1mm]
         Vorobievy Gory, Moscow, 119899, RUSSIA}\\[2mm]
      {\normalsize \bf  E-mail: svernov@theory.npi.msu.su}\\[15mm]
\end{center}
\begin{abstract}
 \normalsize
  Doubly periodic (periodic both in time and in space) solutions for the
Lagrange--Euler equation of the $(1+1)$--dimensional scalar $\varphi^4$
theory are considered. The nonlinear term is assumed to be small, and the
Poincar\'e--Lindstedt method is used to find asymptotic solutions in the
standing wave form. The principal resonance problem, which arises for zero
mass, is solved if the leading-order term is taken in the form of a Jacobi
elliptic function. It have been proved that the choice of
elliptic cosine with fixed value of module k ($k\approx
0.451075598811$) as the leading-order term puts the principal resonance to
zero and allows us to construct (with accuracy to third order of small
parameter) the asymptotic solution  in the standing wave form.  To obtain
this leading-order term the computer algebra system {\bf REDUCE} have been
used. We have appended the {\bf REDUCE} program to this paper.
\end{abstract}

\newpage

\section{Introduction}
\subsection{Periodic solutions of nonlinear equations}

Recently, periodic solutions of the nonlinear wave equation:
$$
\frac{\partial^2\varphi(x,t)} {\partial
 x^2}-\frac{\partial^2\varphi(x,t)} {\partial t^2}-
g(\varphi)=0,\eqno(1)
$$
 where $g(\varphi)$ denotes a continuous function on $\real$ such that
$g(0)=0$, are discussed intensively.  In the 1970s and 1980s existence
theorems for a wide class of continues functions  $g(\varphi)$ satisfying
definite boundary conditions in $x$ have been proved.

J.~M.~Coron~\cite{31} has proved that the following problem:
$$
\left\{ \begin{array}{lcl}
   \displaystyle\frac{\partial^2\varphi(x,t)} {\partial
 x^2}-\frac{\partial^2\varphi(x,t)} {\partial t^2} -g(\varphi)=0,&
{ } &
 \displaystyle\;\; x,t\in \real,\;\; g(0)=0, \\[3mm]
 \displaystyle\lim\limits_{|x|\rightarrow\infty}\varphi(x,t)=0,& {
} &
 \displaystyle\;\;  x,t\in \real,\\[3mm]
\displaystyle \varphi(x,t+T)=\varphi(x,t),
  & { } &\displaystyle\;\;  x,t,T\in \real,\\
\end{array}
\right.
\eqno(2)
$$
has a $T$-periodic (in time) solutions  only if
$g'(0)>\left(\frac{\textstyle 2\pi}{T}\right)^2$.

In particular, if $g(\varphi)=\sin(\varphi)$, there are no $T$-periodic
solutions when $T<2\pi$ \ (for any period $T>2\pi$ an
explicit solution is known, see e.g. G.~Lamb~\cite{41}). If
$g(\varphi)=-\sin(\varphi)$ or $g(\varphi)=\pm\varphi^3$, then
 there is no nontrivial solution of problem $(2)$ (for any period $T$). To
obtain periodic solutions of massless $\varphi^4$ theory one has to change
the boundary conditions.

 Generalizing the Rabinowitz theorem~\cite{33},  H. Brezis, J. M.  Coron
and L.  Nirenberg~\cite{32}  have proved that for each $T$, which is a
rational multiple of $\pi$, the following problem:
$$
 \left\{
 \begin{array}{lcl}
  \displaystyle\frac{\partial^2\varphi(x,t)} {\partial
 x^2}-\frac{\partial^2\varphi(x,t)} {\partial t^2} -g(\varphi)=0,&
{ } &
  \displaystyle x,t\in \real,\;\; g(0)=0,\;\; \\[3mm]
   \varphi(0,t)=\varphi(\pi,t)=0,& {
  } & \displaystyle t\in \real,\\[3mm] \displaystyle
  \varphi(x,t+T)=\varphi(x,t), & { } & \displaystyle x,t,T\in
\real,\\
 \end{array}
\right. \eqno (2')
 $$
has a nontrivial weak $T$-periodic solution if  $g(\varphi)$ is a
continuous nondecreasing function, satisfying the following conditions:
$$
 \lim\limits_{|\varphi|\rightarrow\infty}\frac{g(\varphi)}{\varphi}
 =\infty\;,
$$
$$
\exists \; \alpha,\:\beta>0 \mbox{\hspace{1cm}such
\  that\hspace{1cm}}\forall \varphi\in\real \;{ }\; :  \;{ }\;
\frac{1}{2}\varphi\!\cdot\!  g(\varphi)-\int\limits_0^\varphi\!
g(\psi)\,d\psi\ge \beta|g(\varphi)|-\alpha\;.
$$

By a {\it weak solution} we mean  such  function $\varphi$ that
$$
 \int\limits_0^T\!\int\limits_0^\pi\!\Bigl(\varphi(v_{tt}-v_{xx})+
g(\varphi)v\Bigr)\:dt\:dx\;=\;0
$$
 for all $v\in C^2([0,\pi]\times \real)$ satisfying the boundary and
periodicity conditions. A simplified proof of the Rabinowitz theorem
for the case
$$
g(\varphi)=|\varphi|^n\!\cdot\!\varphi,
$$
where $n>0$, was given in~\cite{34}.

\subsection{Two kinds of doubly periodic solutions}

 Our investigation is dedicated to the construction of doubly periodic
 classical fields in the $(1+1)$--dimen\-sional $\varphi^4$ theory.  We
study the model of an isolated real scalar field $\varphi(x,t)$, described
by the Lagrangian density:

$$
  {\cal L}(\varphi) =
  \frac{1}{2}\left(\varphi_{,t}^2(x,t)-\varphi_{,x}^2(x,t)-
  M^2\varphi^2(x,t) - \frac{\varepsilon}{2}\varphi^4(x,t)\right).
$$

The corresponding Lagrange--Euler equation is:
$$
 \frac{\partial^2\varphi(x,t)}{\partial
 x^2}-\frac{\partial^2\varphi(x,t)} {\partial t^2}-M^2\varphi(x,t)
-\varepsilon\varphi^3(x,t)=0.      \eqno(3)
$$

 There are two classes of doubly periodic  solutions for this equation.
If we  seek fields in the traveling wave form:
$$
\varphi(x,t)=\varphi(x-vt),
$$
where $v$ is the velocity of the wave motion, then equation~$(3)$ reduces
to the Duffing's equation~\cite{5}.  As is known~\cite{2}, periodic
solutions for this equation are Jacobi elliptic functions.  Such
solutions are well known for both $(1+1)$--dimensional \cite{15} and
$(3+1)$--di\-mensional~\cite{46} variants of $\varphi^4$ theory. They
are used in the Yang--Mills theory~\cite{77}.

The second class of doubly periodic solutions consists of functions in
the standing wave form:
$$
  \varphi(x,t)\equiv\sum_{n=1}^{\infty}\sum_{j=1}^{\infty}
{\cal C}_{nj}^{\vphantom{+}}\sin(n(x-x_0^{\vphantom{+}}))
\sin(j\,\omega\,(t-t_0^{\vphantom{+}})),\eqno(4)
$$
where $x_0^{\vphantom{+}}$ and $t_0^{\vphantom{+}}$ are constants
determined by boundary and initial conditions.  Equation $(3)$ is a
translation-invariant, so we can restrict our consideration to the case of
zero $x_0^{\vphantom{+}}$ and $t_0^{\vphantom{+}}$  without~loss of
generality. The choice $x_0^{\vphantom{+}}=0$ corresponds to the boundary
 conditions of problem $(2')$. The period of $\varphi(x,t)$ in $x$ is
 taken to be $2\pi$ and we are interesting in finding the period in $t$.

 Exact standing wave solutions are not known. Approximate solutions for
equation~$(3)$ with $\varepsilon=1$ and $M=1$ have been found under the
assumption that all highest harmonics are zeros:  $\forall n,j > N $ :
${\cal C}_{nj}=0$, where $N$ is a large number. Numerical calculations
have been made~\cite{8} for several values of $N$. The obtained solutions
are very close to each other, but convergence of the sequence of these
solutions as $N$ tends to infinity cannot be proved in numerical framework.

\section { Asymptotic solutions in the standing wave form}

Let us consider periodic solutions for equation~$(3)$, provided that
$\varepsilon$ is small or, equivalently, that these solutions have small
amplitudes.

An asymptotic expansion, containing only bounded functions, is called {\it
a uniform expansion}.  The possibility to obtain a uniform expansion,
using standard asymptotic methods, for example,
Poincar\'e--Lindstedt~\cite{10}  or Krylov--Bogoliu\-bov~\cite{11} methods,
depends on values of the frequencies in the zero approximation
(for a review of such methods, see~\cite{50,550}).  If
$\varepsilon=0$, then equation~$(3)$ is linear and has periodic solutions
of the form~$(4)$ with frequencies in time $\Omega_j=\sqrt{j^2+M^2}$,
where $j \in \natur$.  There are two fundamentally different cases.

If {\LARGE \vspace{-1mm} 
$$ 
{\scriptstyle \forall  i,j \scriptscriptstyle
\in\;\natu} \mbox{ \ \  \  } {\not\:\scriptstyle\exists \; k,n
\scriptscriptstyle\in \;\natu} \mbox{ \ \normalsize  such that \ } {
\scriptstyle\frac{\Omega_{i\vphantom{j}}}{\vphantom{\Omega^+}\Omega_j}
=\frac{k}{n}} \mbox{ \normalsize , } 
$$ 
\normalsize then} it is a nonresonance
case and a uniform expansion can be  found with any desired accuracy using the
standard asymptotic methods, for instance, the Poincar\'e--Lindstedt method.
All secular terms can be eliminated with the proper choice of frequency.

The resonance case, when there exist two frequencies $\Omega_j$ and $\Omega_i$,
whose ratio is rational, is more difficult. The Krylov--Bogoliubov method and
the standard variant of the  Poincar\'e--Lindstedt method can be used to find
periodic solutions only to a few leading orders in~$\varepsilon$.

The important example of resonance case is the massless $\varphi^4$ theory when
$\Omega_j=j$ and all relations of frequencies are rational,  which is the
principal resonance case.  Using standard asymptotic methods, one cannot
construct a periodic solution even to the first order in $\varepsilon$.  It is
possible to transform the differential equations to a system of nonlinear
algebraic equations in Fourier coefficients and frequencies using the
Poincar\'e--Dulac's normal form method.  The algorithm of this procedure was
constructed~\cite{9,99} and realized in the software for symbolic and algebraic
computation {\bf REDUCE}~[18--20] (the system {\bf NORT}~\cite{64,65}).  But an
algorithm to solve the resulting algebraic system has yet to be created.

 \section{ The standing wave solutions for the massless
$\varphi^4$ theory }

 \subsection { Construction of asymptotic solutions via the
Poincar\'e--Lindstedt method }

 The purpose of this article is the construction of standing wave
 solutions of equation~$(3)$ with $M=0$, using asymptotic methods when
$\varepsilon\ll 1$.  We use the Poincar\'e--Lindstedt method: introduce
the new time $\tilde t\equiv \omega t$ and look for a  doubly periodic
 function $\varphi(x,\tilde t)$ and a
 frequency (in time) $\omega$ in the form of power series in
$\varepsilon$:
 $$
 \begin{array}{r@{\;\equiv\;}l}
  \displaystyle \varphi(x,\tilde t,\varepsilon) &\displaystyle
\sum_{n=0}^\infty\varphi_n^{\vphantom{+}} (x,\tilde
t)\varepsilon^n,
\\[0.4cm] \displaystyle \omega(\varepsilon)& \displaystyle
1+\sum_{n=1}^\infty\omega_n^{\vphantom{+}} \varepsilon^n.
\end{array}
$$

 Expanding the Lagrange--Euler equation in a  power series in
$\varepsilon$, we obtain a series of equations. The first two
equations of this series are:
\begin{itemize}
\item to zero order in $\varepsilon$ the equation for
  $\varphi_0^{\vphantom{+}}$ is:
$$
  \frac{\partial^2\varphi_0^{\vphantom{+}}(x,\tilde t)}{\partial
x^2}-
\frac{\partial^2\varphi_0^{\vphantom{+}}(x,\tilde t)}
{\partial \tilde t^2}=0,   \eqno(5)
$$

\item to first order in $\varepsilon$ the equation in
  $\varphi_0^{\vphantom{+}}$,
$\varphi_1^{\vphantom{+}}$ and  $\omega_1^{\vphantom{+}}$ is:
 $$ \frac{\partial^2\varphi_1^{\vphantom{+}}(x,\tilde t)}{\partial x^2}-
\frac{\partial^2\varphi_1^{\vphantom{+}}(x,\tilde t)}{\partial \tilde t^2}
=2\omega_1^{\vphantom{+}} \frac{\partial^2\varphi_0(x,\tilde t)}{\partial
\tilde t^2} +\varphi_0^3(x,\tilde t).  \eqno(6)
$$

\end{itemize}

  Equation $(5)$ has many periodic solutions.  If we select
$\varphi_0^{\vphantom{+}}(x,\tilde t)=\sin(x)\sin(\tilde t)$,  then the
second equation has no periodic solution, because the frequency  of the
external force $\sin(3x)\sin(3\tilde t)$ is equal to the frequency of its
own oscillations and it is impossible to put this resonance harmonic to
zero by selecting only $\omega_1^{\vphantom{+}}$. In the massless case to
obtain a uniform expansion, we must find another doubly periodic solution
to equation $(5)$. The correct selection of not only the frequency
$\omega(\varepsilon)$, but also the function
$\varphi_0^{\vphantom{+}}(x,\tilde t)$, allows us to construct the next to
leading order term in standing wave form.

\subsection{ The condition of existence of periodic solutions }

   The general solution, in the standing wave form $(4)$, for
equation~$(5)$ is the function
 $$
 \varphi_0^{\vphantom{+}}(x,\tilde t)=
 \sum_{n=1}^{\infty}a_n^{\vphantom{+}}\sin(nx)\sin(n\tilde t)
$$
 with arbitrary $a_n^{\vphantom{+}}$.  We have to find coefficients
$a_n^{\vphantom{+}}$ so that the function
$\varphi_1^{\vphantom{+}}(x,\tilde t)$ is a periodic solution for
 equation~$(6)$.  If we select $\varphi_1^{\vphantom{+}}(x,\tilde t)$ as a
double sum:
$$
\varphi_1^{\vphantom{+}}(x,\tilde
t)\equiv\sum_{n=1}^{\infty}\sum_{j=1}^{\infty}
b_{nj}^{\vphantom{+}}\sin(nx)\sin(j\tilde t)
$$
with arbitrary $b_{nj}^{\vphantom{+}}$, then equation~$(6)$ can be
presented in the form of Fourier series:
$$
\frac{\partial^2\varphi_1^{\vphantom{+}}(x,\tilde t)}{\partial
 x^2}  -\frac{\partial^2\varphi_1^{\vphantom{+}}(x,\tilde
t)}{\partial
\tilde t^2} -2\omega_1^{\vphantom{+}}
\frac{\partial^2\varphi_0^{\vphantom{+}}(x,\tilde
t)}{\partial \tilde t^2}
-\varphi_0^3(x,\tilde t)=\sum_{n=1}^{\infty}\sum_{j=1}^{\infty}
R_{nj}({\bf a},{\bf b})\sin(nx)\sin(j\tilde t)=0
$$
 and is equivalent to the following infinite system of the algebraic
equations in Fourier coefficients of the functions
$\varphi_0^{\vphantom{+}}(x,\tilde t)$ and
 $\varphi_1^{\vphantom{+}}(x,\tilde t)$:
 $$
 \forall n,j \;{}:{}\;  R_{nj}({\bf a},{\bf b})=0. \eqno(6')
 $$

This system has a subsystem of the equations in the Fourier coefficients
 of the function $\varphi_0^{\vphantom{+}}(x,\tilde t)$:
$$
\forall j\in\natu\;{ }:{ }\;
  R_{jj}({\bf a})=0, \eqno(7) $$ where $$ \begin{array}{rcl}
\displaystyle
R_{jj}&\displaystyle \!\!\equiv\!\! & \displaystyle
9a_j^3+3a_j^2a_{3j}^{\vphantom{+}} +a_j^{\vphantom{+}}\!\left(6\!
 \sum_{\vbox{\hbox{$\scriptstyle
s=1$}\vspace{1mm}\hbox{$\scriptstyle
s\neq j$}}}^\infty\;(2a_s^2+
a_s^{\vphantom{+}}a_{2j+s}^{\vphantom{+}})+
 3\!\sum_{\vbox{\hbox{$\scriptstyle s=1$}
\vspace{1mm}\hbox{$\scriptstyle s\neq j$}}}^{2j-1}\!
a_s^{\vphantom{+}}a_{2j-s}^{\vphantom{+}}-
 32j^2\omega_1^{\vphantom{+}}\right)+\\[1cm]
&\displaystyle \!\!+\!\!&\displaystyle
3\sum_{\vbox{\hbox{$\scriptstyle s=1$}
\vspace{1mm}\hbox{$\scriptstyle s\neq j$}}}^{\infty}
\;\sum_{\vbox{\hbox{$\scriptstyle
p=1$}\vspace{0.5mm}\hbox{$\scriptstyle
p\neq j$}}} ^\infty a_s^{\vphantom{+}} a_p^{\vphantom{+}}
a_{j+s+p}^{\vphantom{+}} +3 \sum_{\vbox{\hbox{$\scriptstyle s=1$}
\vspace{1mm}\hbox{$\scriptstyle s\neq j$}}}^{\infty}\!\!\!\!\!\!
\sum_{\vbox{\hbox{\ \ \ \ $\scriptstyle p=1$}\vspace{0.5mm}\hbox{\
\ \ \
$\scriptstyle p\neq j$}\vspace{1mm}\hbox{\ \ \ \ $\scriptstyle
p\neq
2j-s$}}}^{\infty}\!\!\!\!\!\!\!a_s^{\vphantom{+}}
a_p^{\vphantom{+}}a_{s+p-j}^{\vphantom{+}}+
\sum_{s=1}^{j-2}\;\sum_{p=1}^{j-2}
a_{s}^{\vphantom{+}}a_{p}^{\vphantom{+}}a_{j-s-p}^{\vphantom{+}}
\end{array}
$$

If the system~$(7)$ is solved and all Fourier coefficients of the function
$\varphi_0^{\vphantom{+}}(x,\tilde t)$ are known, then the system~$(6')$
transforms into a system of linear equations: $R_{nj}({\bf b})=0$, which
always has only one solution. Hence, we have obtained the necessary and
sufficient condition of the existence of periodic solutions for
equation~$(6)$: there exists a periodic function
$\varphi_1^{\vphantom{+}}(x,\tilde t)$, satisfying equation~$(6)$, if and
only if the Fourier coefficients of the function
$\varphi_0^{\vphantom{+}}(x,\tilde t)$ satisfy system~$(7)$.

The coefficient $a_1^{\vphantom{+}}$ is a parameter determining the
oscillation amplitude. In fact, let
$a_j^{\vphantom{+}}=c_j^{\vphantom{+}}a_1^{\vphantom{+}}$ \  and \
$\omega_1^{\vphantom{+}}=c_\omega^{\vphantom{+}}a_1^2$, then all
polynomials $R_{jj}({\bf a})$ are proportional to $a_1^3$:  $R_{jj}({\bf
a})=a_1^3R_{jj}({\bf c})$ and, therefore, the coefficient
$a_1^{\vphantom{+}}$ can be selected arbitrarily.

\subsection{ The approximate solution of system~(7)}

 System $(7)$ is very difficult to solve. On the one hand, all $R_{jj}$
are infinite series and the number of equations is infinite too.  On the
 other hand, each equation of this system is nonlinear. We restrict
 ourselves to finding a particular solution. To simplify calculations, we
 assume that the function $\varphi_0^{\vphantom{+}}(x,\tilde t)$ contains
only odd harmonics. Our goal is to find a real solution so we seek
$c_j^{\vphantom{+}}\in\real$.

To find an approximate solution we apply the Galerkin method:
cut off higher diagonal harmonics $\forall j>N$:
$c_{2j-1}^{\vphantom{+}}=0$ and seek an approximation for
 $\varphi_0^{\vphantom{+}}(x,\tilde t)$ in the following form
$$
\varphi_0^{\vphantom{+}}(x,\tilde t)=a_{1}^{\vphantom{+}}
  \left\{\sum_{j=1}^{N}
c_{2j-1}^{\vphantom{+}}\sin((2j-1)x)\sin((2j-1)\tilde t)\right\}.
$$
Neglecting the higher harmonics of the
function $\varphi_0^{\vphantom{+}}(x,\tilde t)$, we obtain a finite
system of nonlinear equations, with the number of equations three times as
many as that of the variables.  We solve the system of the $N$ leading
equations and substitute the obtained values in the other equations. In
principle, any system of $N$ nonlinear equations in $N$ unknowns can be
solved with a computer, using Buchberger algorithm~\cite{123,124}.  This
algorithm has been realized in the standard procedures of the computer
algebra system {\bf REDUCE}: the {\bf SOLVE} operator and the Gr\"oebner
basis package.  This algorithm allows us to diagonalize this system of
nonlinear equations, constructing an equivalent system, which consists of
equation in only one variable, equation in two variables and so on. So, we
can obtain solution of the system, solving only equations in one variable
and substitute the result into the remaining equations.

The operating memory and other parameters of real computers are not enough
to solve very difficult systems.  For example, using computer with 128
Mbytes operating memory (RAM), we have solved the system, consisting of
the $N$ leading equations of the system~$(7)$ only for $N<6$. On the other
hand, even for these values of $N$ the solution of system can be
found only with finite accuracy (the exact solution can be irrational).

The {\bf SOLVE} operator in {\bf REDUCE} (in "on rounded" mode) rounds off
numbers with accuracy $\delta=10^{-11}$. We assume that the system of the
$N$ leading equations is solved if for all $j\leq N$ the inequalities
$|R_{jj}({\bf c})|< \delta$ are true. We also admit that the value of $N$
is sufficient to solve system~$(7)$ if for all $j\in \natur$ the
inequalities $|R_{jj}({\bf c})|< \delta$ are true.

To find a particular solution of the $N$ leading equations for any $N<50$
we have written a short program in {\bf REDUCE } (see {\it Appendix}) and
obtained that the minimal sufficient value of $N$ for $\delta=10^{-11}$ is
$N=8$ and that the frequency correction is
$$
\omega_1^{\vphantom{+}}=0.28268003454\:a_1^2.
$$

We also have found numerical
values of the fifteen leading Fourier coefficients of
$\varphi_0^{\vphantom{+}}(x,\tilde t)$. The values of $c_j^{\vphantom{+}}$ are
very close to the values of the corresponding terms of the following finite
sequence: $$ {\bf d}=\{ d_{2j-1}^{\vphantom{+}}=\frac{f_{2j-1}^{\vphantom{+}}}
{f_1^{\vphantom{+}}}\mbox{, \ where \  \  } { }
f_{2j-1}^{\vphantom{+}}\equiv\frac{q^{j-1/2}}{1+q^{2j-1}}; {} \;\;{ }
d_{2j}^{\vphantom{+}}=0 ; { } \ \;{ } j<23 \}, $$ with $q=0.0142142623201$. It
is easy to verify that substitution of this finite sequence gives $ \forall
j\in\natur \mbox{ \ \ : \ \ }|R_{jj}({\bf d})|<10^{-12}$.

In other words, the finite sequence  ${\bf d}$
is an approximate solution of system~$(7)$.
These numerical calculations help to find the analytical form of
$\varphi_0^{\vphantom{+}}(x,\tilde t)$. The following
table illustrates this interesting result:

\begin{center}
\begin{tabular}{|c|c|c|c|c|}
\hline
$j^{\vphantom{+}}$ & $c_{j_{\vphantom{j}}}^{\vphantom{+}}$ &
$R_{jj}({\bf
c})$ & $d_{j_{\vphantom{j}}}^{\vphantom{+}}$ & $R_{jj}({\bf d})$
\\ \hline

1     & 1     &  0  &  $1^{\vphantom{+}}$ & 0 \\ \hline

3 &$\;1.44162661711\times 10^{-2^{\vphantom{+}}}\;$  &   $
3.5\times
10^{-12}$ & $\;1.44162661711\times 10^{-2^{\vphantom{+}}}\;$     &
$ -
8.0\times 10^{-14}$ \\ \hline

5 &$2.04917177408\times 10^{-4^{\vphantom{+}}}$  &  $ 2.1\times
10^{-12}$
& $2.04917177419\times 10^{-4}$     & $ - 3.4\times 10^{-15}$
\\ \hline

7 & $2.91274649724\times 10^{-6^{\vphantom{+}}}$ &  $ 7.8\times
10^{-12}$
& $2.91274651543\times 10^{-6}$    &  $- 9.7\times 10^{-17}$  \\
\hline

9 & $4.14025418115\times 10^{-8^{\vphantom{+}}}$ &  $ 8.3\times
10^{-13}$
& $4.14025430425\times 10^{-8}$  &   $- 2.3\times 10^{-18}$ \\
\hline

11& $5.88506592014\times 10^{-10^{\vphantom{+}}}$ &  $ 1.0\times
10^{-14}$ & $5.88506607528\times 10^{-10}$&      $- 4.9\times
10^{-20}$
\\ \hline

13& $8.36488192079\times 10^{-12^{\vphantom{+}}}$ & $ 4.6\times
10^{-13}$
& $8.36518729655\times 10^{-12}$       &   $- 9.7\times 10^{-22}$

\\ \hline

15& $1.18901919266\times 10^{-13^{\vphantom{+}}}$ &  $-7.8\times
10^{-22}$  & $1.1890496659\times 10^{-13}$       & $- 1.8\times
10^{-23}$
\\ \hline

17&  0   &  $4.4\times 10^{-12^{\vphantom{+}}}$ &
$1.69014638629\times
10^{-15}$ & $- 3.4\times 10^{-25}$       \\ \hline

19& 0 &  $7.3\times 10^{-14^{\vphantom{+}}}$ &
$2.40241840942\times
10^{-17}$ & $- 6.0\times 10^{-27}$         \\ \hline

21& 0 &   $1.1\times 10^{-15^{\vphantom{+}}}$ &
$3.41486054743\times
10^{-19}$ & $ - 1.0\times 10^{-28}$      \\ \hline

23& 0 &   $1.7\times 10^{-17^{\vphantom{+}}}$  &
$4.85397236079\times
10^{-21}$ &  $   - 1.8\times 10^{-30}$       \\ \hline

25&   0 &  $2.4\times 10^{-19^{\vphantom{+}}}$ &   $
6.89956364312\times 10^{-23}$ &   $ - 3.0\times 10^{-32}$      \\
\hline

27&  0 &   $3.3\times 10^{-21^{\vphantom{+}}}$ &    $
9.8072207518\times
10^{-25}$ &   $  - 4.9\times 10^{-34}$             \\ \hline

29&  0 &   $4.2\times 10^{-23^{\vphantom{+}}}$  &   $
1.39402408398\times
10^{-26}$       & $ - 8.1\times 10^{-36}$       \\
\hline

31&   0 &  $5.2\times 10^{-25^{\vphantom{+}}}$  &     $
     1.98150240103\times 10^{-28}$ & $- 1.3\times 10^{-37}$
\\
\hline

33&    0 &  $5.7\times 10^{-27^{\vphantom{+}}}$ &      $
2.81655949163\times 10^{-30}$ & $- 2.1\times 10^{-39}$
\\
     \hline

35&    0  &  $6.1\times 10^{-29^{\vphantom{+}}}$  &      $
4.00353154544\times 10^{-32}$ &   $- 3.4\times 10^{-41}$        \\
\hline

37& 0  &  $6.2\times 10^{-31^{\vphantom{+}}}$     &
$5.69072475939\times
10^{-34}$         &  $ - 5.4\times 10^{-43^{\vphantom{+}}}$
\\ \hline

39& 0  &  $5.9\times 10^{-33}$  &   $8.08894545219\times
10^{-36}$   &  $ - 8.5\times 10^{-45^{\vphantom{+}}}$
\\
\hline

41& 0  &  $5.0\times 10^{-35}$  &  $ 1.14978392551\times 10^{-37}$
&  $- 1.3\times 10^{-46^{\vphantom{+}}}$           \\ \hline

43&  0  &  $ 3.6\times 10^{-37}$   &   $ 1.63433303287\times
10^{-39}$  & $ - 2.1\times 10^{-48^{\vphantom{+}}}$           \\
\hline

45&  0  &  $ 1.7\times 10^{-39^{\vphantom{+}}}$  &   $
2.32308384477\times 10^{-41}$ &$- 1.6\times 10^{-
49^{\vphantom{+}}}$\\\hline
47& 0 & 0 & 0 &$1.4\times 10^{-50^{\vphantom{+}}}$ \\\hline
\end{tabular}
\end{center}

\subsection { The exact solution of system~(7)}

The approximate solution of system $(7)$ is found.
The aim of this section is to find an exact solution of $(7)$,
that is to say to find the function $\varphi_0^{\vphantom{+}}(x,
t)$ in analytical form.
 For arbitrary $q\in(0,1)$ let us define the following
sequence:
$$
  {\bf f}\stackrel{\rm def}{=}\{\;\forall n\in
  \natu { }:{ } f_{2n-1}^{\vphantom{+}}=\frac{q^{n-1/2}}{1+q^{2n-
1}},{ }
  {    }{ } f_{2n}^{\vphantom{+}}=0\;\}.
$$
 The terms of the sequence ${\bf f}$ are proportional to Fourier
coefficients of the Jacobi elliptic function $\bf cn$ 
(the elliptic cosine):
$$
\cn(z,k)=\frac{\gamma}{k}\sum_{n=1}^\infty
f_{2n-1}^{\vphantom{+}}\cos\left((2n-1)\frac{\gamma z}{4} \right)
\mbox{, \ \ where \ } \gamma\equiv\frac{2\pi}{K} \mbox{ , \  \  \
}
z\in \real.
\eqno (8)
$$

Let us clarify the notation and point out some properties of
the elliptic cosine~\cite{2}:

\begin{itemize}

\item Basic periods of the doubly periodic function $\cn(z,k)$ are
$4K(k)$ and $2K(k)+2\dot\imath K'(k)$, where $K(k)$ is a full elliptic
integral, $K'(k)\equiv K(k')$ and $k'=\sqrt{1-
k^{2^{\vphantom{+}}}}$.\vspace{1mm}

\item The parameter $q$ in the Fourier expansion can be expressed in terms
of elliptic integrals:  $q\equiv e^{-
{\displaystyle\pi}\frac{K'}{K^{\vphantom{+}}}}$.\vspace{1mm}

\item The Fourier-series expansion of the function $\cn(z,k)$ does
not include even harmonics.  This expansion is valid in the
following domain of the complex plane: $-K'<\Im m\: z<K'$, in particular,
for $z\in \real$.  \vspace{1mm}

\item If $z \in \real$ and $k\in(0,1)$, then  $\cn(z,k)\in \real$.
\vspace{1mm}

\item The function $\cn(z,k)$ is a solution of the following
 differential equation:
 $$
 \frac{d^2\cn(z,k)}{dz^2}=(2k^2-1)\cn(z,k)-2k^2\cn^3(z,k).
 \eqno (9)
 $$
\end{itemize}

The latest property means that the infinite sequence of the Fourier
coefficients for $\cn(z,k)$ is a solution of some infinite system of
nonlinear algebraic equations. Let us find this system.  On the one
hand, it is clear from $(8)$ that the Fourier-series expansion for the
function $\cn^3(z,k)$ is:

$$
\cn^3(z,k)=\frac{\gamma^3}{4k^3}\sum_{j=1}^\infty F_j^{(3)}({\bf
f})
\cos\left(j\frac{\gamma z}{4}\right)\mbox{, \ \ where }
j=1,3,5,\dots,+\infty
\mbox{\ ; \ \ \ }
$$
$$
\begin{array}{rl}
\displaystyle
F_{j}^{(3)}({\bf f})\!&\displaystyle=\;
3f_j^3+3f_j^2f_{3j}^{\vphantom{+}} +f_j^{\vphantom{+}}\left(6\sum_
{\vbox{\hbox{$\scriptstyle s=1$}
\vspace{1mm}\hbox{$\scriptstyle  s\neq j$}}}^{\infty}\;(f_s^2+
f_s^{\vphantom{+}}f_{2j+s}^{\vphantom{+}})+
 3\!\sum_{\vbox{\hbox{$\scriptstyle s=1$}
\vspace{1mm}\hbox{$\scriptstyle s\neq j$}}}^{2j-1}
 f_s^{\vphantom{+}}f_{2j-s}^{\vphantom{+}} \right)+\\[1cm]
\!&\displaystyle+\;\;
3\sum_{\vbox{\hbox{$\scriptstyle s=1$}
\vspace{1mm}\hbox{$\scriptstyle s\neq
j$}}}^{\infty}\;\sum_{\vbox{\hbox{$\scriptstyle p=1$}
\vspace{0.5mm}\hbox{$\scriptstyle p\neq
j$}}}^{\infty}f_s^{\vphantom{+}}
f_p^{\vphantom{+}} f_{j+s+p}^{\vphantom{+}} +3
\sum_{\vbox{\hbox{$\scriptstyle s=1$}
\vspace{1mm}\hbox{$\scriptstyle s\neq
j$}}}^{\infty}\!\!\!\!\!\!\!\sum_{\vbox{\hbox{\ \  \ \
$\scriptstyle p=1$}\vspace{0.5mm}\hbox{\ \ \ \ $\scriptstyle p\neq
j$}\vspace{1mm}\hbox{\ \ \ \ $\scriptstyle p\neq 2j-s$}}}^{\infty}
\!\!\!\!\!\!\!\! f_s^{\vphantom{+}}
f_p^{\vphantom{+}}f_{s+p-j}^{\vphantom{+}}+
\sum_{s=1}^{j-2}\;\sum_{p=1}^{j-2}
f_{s}^{\vphantom{+}}f_{p}^{\vphantom{+}}f_{j-s-p}^{\vphantom{+}}
\end{array}
$$
(in all sums we summarize over only odd numbers).

On the other hand, from the differential equation~$(9)$ it follows that
$F_j^{(3)}({\bf f})$ is proportional to $f_j^{\vphantom{+}}$, with
coefficients of proportionality depending on $j$:
$$
\forall j \mbox{ \ \
:  \ \ } F_j^{(3)}({\bf
f})=\left(\frac{2(2k^2-1)}{\gamma^2}+\frac{j^2}{8}\right)
f_j^{\vphantom{+}}.    \eqno(10)
$$

 Thus, the sequence ${\bf f}$ is a nonzero solution of system~$(10)$ at
all $q\in(0,1)$.  The following lemma proves the existence of
the selected value of $q$.

\vspace{5mm}
\noindent $\!\!\!\!\!$
\begin{tabular}{rp{12.4cm}}
\large\bf Lemma. & \large\it  There exists such value of parameter
$q\in(0,1)$ that the sequence ${\bf f}$ becomes a  real solution to
system~$(7)$ with the real-valued parameter \ $\omega_1^{\vphantom{+}}$.
\\[1mm]
 \bf Proof. &  Inserting the sequence ${\bf f}$ into system  $(7)$ :
$a_j^{\vphantom{+}}=f_j^{\vphantom{+}}$ and using system  $(10)$, we
 obtain:

\end{tabular}

 $$
 \begin{array}{r@{\displaystyle \;=\;}l}
\displaystyle \mbox{ \ \  }
\forall j \mbox{ \ \   :  \ \ }
 R_{jj}({\bf f})& \displaystyle \left\{ F_j^{(3)}({\bf f})+
 f_j^{\vphantom{+}}\left(6\sum_{n=1}^\infty f_n^2
 -32j^2\omega_1^{\vphantom{+}}
  \right)\right\}=\\[0.7cm]
 &\displaystyle f_j^{\vphantom{+}}
 \left\{6\sum_{n=1}^\infty f_n^2+
 \frac{2(2k^2-1)}{\gamma^2}+j^2\left(\frac{1}{8}-32
 \omega_1^{\vphantom{+}}\right)\right\}=0.
\end{array}
 $$

 System  $(7)$ has a nonzero solution if and only if

$$
 \left\{ \begin{array}{r@{\; =\;}l}
 \displaystyle\omega_1^{\vphantom{+}}&\displaystyle
 \frac{1}{256},\\[0.3cm]
 \displaystyle\sum\limits_{n=1}^\infty
    f_n^2 & \displaystyle\frac{(1-2k^2)}{3\gamma^2}. \\
\end{array}
\right.
$$

 We have obtained the value of $\omega_1^{\vphantom{+}}$. The second
equation of this system is equivalent to the following equation in
parameter $q$:

$$
3\sum_{n=1}^\infty\left(\vphantom{\sum_{n=1}^\infty}
\frac{q^{n-1/2}}{1+q^{2n-1}}\right)^2
-\left(\frac{1}{4}+ \sum_{n=1}^\infty
\frac{q^n}{1+q^{2n}}\right)^2+2\left(\sum_{n=1}^\infty
\frac{q^{n-1/2}}{1+q^{2n-1}}\right)^2=0.     \eqno (11)
$$

This equation has the following solution on interval (0,1):

$$ q=1.42142623201\times10^{-2}\pm1\times10^{-13}.  $$

\noindent{\bf Thus \ the \ lemma \ is \ proved.}

 We can now construct the desired leading-order
approximation for the function $\varphi(x,\tilde t)$:
$$
\varphi_0^{\vphantom{+}}(x,\tilde t)= A\Bigl\{\cn(\alpha (x-\tilde
t),k)-\cn(\alpha (x+\tilde t),k)\Bigr\}.
$$

For arbitrary $k\in(0,1)$ this function is a real solution of equation
  $(6)$. If $\alpha=\frac{2K}{\displaystyle\pi}$, then the periods of
  $\varphi_0^{\vphantom{+}}(x,\tilde t)$ in $x$ and in $\tilde t$ are
  equal to $2\pi$. Using the Fourier-series expansion for the function
$\cn(z,k)$ (formula $(8)$), we obtain the following  expansion for
function $\varphi_0^{\vphantom{+}}(x,\tilde t)$:

$$
\varphi_0^{\vphantom{+}}(x,\tilde
t)=2A\frac{\gamma}{k}\sum_{n=1}^{\infty}f_{2n-1}^{\vphantom{+}}
\sin((2n-1)x)\sin((2n-1)\tilde t).
$$

 If $q=1.42142623201\times10^{-2}\pm1\times10^{-13}$, then $q$ is a
solution of equation~$(11)$ and the sequence ${\bf f}$ is a real solution
of system~$(7)$.  The middle value of $q$ corresponds to
$k=0.451075598811$, $\gamma=3.78191440007$ and $\alpha=1.0576653982$.  All
equations in system $(7)$ are homogeneous, hence for  these values of
parameters, the sequence of the Fourier coefficients of the function
$\varphi_0^{\vphantom{+}}(x,\tilde t)$ also is a solution of system~$(7)$,
with
$$
\omega_1^{\vphantom{+}}=\frac{\gamma^2}{64k^2}A^2=1.0983600974\:A^2.
$$

 Thus we have proved that the function

$$
\varphi_0^{\vphantom{+}}(x,\tilde t)= A\Bigl\{\cn(\alpha (x-\tilde
t),k)-\cn(\alpha (x+\tilde t),k)\Bigr\},
$$
with $k=0.451075598811$ and
$\alpha=1.0576653982$ is such standing wave solution of equation~$(5)$  that
equation~$(6)$ has a periodic solution\footnote{This result has been published
in our paper~\cite{27021998}}.

\subsection{The first  approximation}

 Now it is easy to find the periodic solution
 $\varphi_1^{\vphantom{+}}(x,\tilde t)$.  Let us designate the Fourier
coefficients of the function $\varphi_0^3(x,\tilde t)$ as $D_{nj}$:

 $$
\varphi_0^3(x,\tilde t)\equiv\sum_{n=1}^{\infty}
 \:\sum_{j=1}^{\infty}D_{nj}\sin(nx) \sin(j\tilde
 t).
 $$

 Equation $(6)$ gives the following result ($b_{nn}^{\vphantom{+}}$ are
arbitrary numbers):

$$
\varphi_1^{\vphantom{+}}(x,\tilde t)=\sum_{n=1}^\infty
 \:\sum_{\vbox{\hbox{$\scriptstyle j=1$}\vspace{-1mm}
\hbox{$\scriptstyle
 j\neq n$}}}^\infty\frac{D_{nj}}{j^2-n^2}
\sin(nx)\sin(j\tilde t)+
\sum_{n=1}^{\infty}b_{nn}^{\vphantom{+}}\sin(nx)\sin(n\tilde t).
$$

 It should be noted that the function $\varphi_1^{\vphantom{+}}(x,\tilde
 t)$ with arbitrary diagonal coefficients $b_{nn}^{\vphantom{+}}$ is a
solution of equation~$(6)$ and that all off-diagonal coefficients of
$\varphi_1^{\vphantom{+}}(x,\tilde t)$ are proportional to $A^3$.

 \subsection{The second approximation}

 Let us consider the approximation to second order in
$\varepsilon$:

$$
 \frac{\partial^2\varphi_2^{\vphantom{+}}(x,\tilde
   t)}{\partial x^2} -\frac{\partial^2\varphi_2^{\vphantom{+}}(x,
\tilde t)}{\partial \tilde t^2}
=2\omega_1^{\vphantom{+}}
\frac{\partial^2\varphi_1^{\vphantom{+}}(x,\tilde
t)}{\partial\tilde
t^2}+(2\omega_2^{\vphantom{+}}+\omega_1^2)
\frac{\partial^2\varphi_0^{\vphantom{+}}(x,\tilde t)}{\partial
\tilde t^2}
+3\varphi_1^{\vphantom{+}}(x,\tilde t)
\varphi_0^2(x,\tilde t).    \eqno(12)
$$

If  all diagonal
 coefficients of $\varphi_1^{\vphantom{+}}(x,\tilde t)$ are zeros:
 $\forall n$ :\  $b_{nn}^{\vphantom{+}}=0$,
then $\forall j,n$ :\  $b_{jn}^{\vphantom{+}}=-
b_{nj}^{\vphantom{+}}$~, and
 the function  $\varphi_1^{\vphantom{+}}(x,\tilde
t)\varphi_0^2(x,\tilde
  t)$  has no diagonal harmonics. Hence, selecting
$\omega_2^{\vphantom{+}}=-\frac{1}{2}\omega_1^2$~,
we obtain a periodic solution to equation~$(12)$:

$$
\varphi_2^{\vphantom{+}}(x,\tilde t)\equiv\sum_{n=1}^\infty
 \:\sum_{\vbox{\hbox{$\scriptstyle j=1$}\vspace{-1mm}
\hbox{$\scriptstyle
j\neq n$}}}^\infty\frac{H_{nj}}{j^2-n^2}\sin(nx)\sin(j\tilde t)+
\sum_{n=1}^\infty h_{nn}^{\vphantom{+}}\sin(nx)\sin(n\tilde t),
$$
where
$$
 H(x,\tilde t)\equiv
2\omega_1^{\vphantom{+}}\frac{\partial^2
\varphi_1^{\vphantom{+}}(x,\tilde t)}
{\partial \tilde t^2} + 3\varphi_1^{\vphantom{+}}(x,\tilde
t)\varphi_0^2(x,\tilde t)
\equiv\sum_{n=1}^\infty\:\sum_{j=1}^\infty H_{nj}
  \sin(nx)\sin(j\tilde t).
 $$
It should be noted that all diagonal coefficients
$h_{nn}^{\vphantom{+}}$
are arbitrary.

\section{Conclusions}

 Using massless $\varphi^4$ theory as an example, we have shown that a
uniform expansion of solutions for quasi\-linear Klein--Gordon equations
can be constructed even in the principal resonance case.  To construct the
uniform expansion we have used the Poincar\'e--Lindstedt method and the
nontrivial zero approximation: the~function

$$
 \varphi_0^{\vphantom{+}}(x,t)=
 A\Bigl\{\cn(\alpha (x-\omega t),k)-\cn(\alpha (x+\omega
t),k)\Bigr\},
 $$
 with $k=0.451075598811$ and $\alpha=1.0576653982$.

Thus, using the Jacobi elliptic function {\bf cn} instead of the
trigonometric function {\bf cos}, we have put the principal resonance to
zero and constructed, with accuracy ${\cal O}(\varepsilon^3)$, the doubly
periodic solution in the standing wave form:

 $$ \varphi(x,\omega
t)=\varphi_0^{\vphantom{+}}(x,\omega t)+
\varepsilon\varphi_1^{\vphantom{+}}(x,\omega t)+
\varepsilon^2\varphi_2^{\vphantom{+}}(x,\omega t)+
{\cal O}(\varepsilon^3),
$$
with the frequency
$$
 \omega=1+\frac{\gamma^2}{64k^2}A^2
\varepsilon-\frac{\gamma^4}{8192k^4}A^4
 \varepsilon^2+{\cal O}(\varepsilon^3)=1+1.0983600974\:
A^2\varepsilon
 -0.6031974518\: A^4\varepsilon^2 +{\cal O}(\varepsilon^3).
$$

The question of constructing the uniform asymptotic series on the base of
the obtained approximation remains open. The authors plan to investigate
this question in future publications.

\section*{ Acknowledgement }

  The authors are grateful to\ \  M.~V.~Tchitchikina,\,\,
V.~F.~Edneral\,\, and\,\,  P.~K.~Silaev\,\,  for valuable discussions.
S.~Yu.~Vernov would like to thank the Organizing Committee of the Fourth
International IMACS Conference on Applications of Computer Algebra
(ACA'98) for invitation to the Conference, hospitality and financial
support.

 This work has been supported by the Russian Foundation for Basic Research
 under Grants {\the\textfont2
N}\lower-0.4ex\hbox{\small\underline{$\circ$}}~96-15-96674 and
{\the\textfont2 N}
\lower-0.4ex\hbox{\small\underline{$\circ$}}~98-02-27299.

\section*{Appendix}

This program have been written in {\bf REDUCE 3.6}.

The program constructs system~$(7)$:  $R_{jj}=0$ and finds a particular
solution of this system.  The function $\varphi_0^{\vphantom{+}}(x,\tilde
t)$ has been denoted as phi(xx,tt). This function is a solution of
equation~$(5)$. We assume that the function phi(xx,tt) is real and
includes only odd harmonics:

$$
phi(xx,tt)=\sum_{j=1}^{n}
a(2j-1)^{\vphantom{+}}\sin((2j-1)xx)\sin((2j-1)tt).
$$

The number of unknowns (parameter n) can be selected arbitrary.
For $n=10$ this program gives result on computer with 16 Mbytes
operating memory, for $n=20$ this program gives result on computer
with 128 Mbytes operating memory.

Using the procedure "fourier" we expand equation~$(6)$ in the Fourier
series and obtain the list of equations "listequa". The number of
equations is $3n$.  The unknowns $a\:(j)$ are a Fourier coefficients of
some function, hence, we can assume that the sequence of the numbers
$c\:(j)\equiv a\:(j)/a\:(1)$, satisfies the following condition: $\forall
j>1$ : $|c\:(j)|<1$.

At the first step we assume that for $j>3$, all $c\:(j)=0$. From the first
equation of "listequa" we find that $C_{omega}$ is a polynomial in
$c\:(3)$. We substitute this value of $C_{omega}$ in the second equation
of "listequa". As result we obtain that this equation is a cubic equation
in $c\:(3)$. We solve this equation, using the standard procedure {\bf
SOLVE}. One of its solution must be real and we select this solution as
value of $c\:(3)$.

At the second step we assume that $c\:(5)$ is arbitrary \{ clear $c\:(5)$
\}.  From the first equation of "listequa" we obtain that $C_{omega}$ is a
polynomial in $c\:(5)$, substitute this value of $C_{omega}$ in the third
equation of "listequa" and solve it, determining $c\:(5)$.

At the third step  we substitute new values of $c\:(5)$ and $C_{omega}$ in
the second equation of "listequa", so we calculate the variable "equa". If
\ "equa" is more than $\delta=10^{-11}$, then we resolve the second
equation of "listequa" and repeat the second step. Alternatively, if the
obtained value of "equa" is less than or equal to $\delta=10^{-11}$, then
we assume that the three leading equations of "listequa" are solved and
seek $c\:(7)$ from the fourth equation of "listequa" and so on.  Once the
coefficient $c\:(2n-1)$ has been found the program finishes to work.

\begin{verbatim}


out "rjj.res";
in fourier$
n:=10$
operator a,c,phi$
depend phi,xx,tt$

% Expand the unknown function phi(xx,tt), which is a solution of
% equation (5), and equation (6) in the Fourier series.

phi(xx,tt):=for j:=1:n sum a(2j-1)*sin((2j-1)*xx)*sin((2*j-1)*tt)$
equation1:=2*w1*df(phi(xx,tt),tt,2)+
                fourier(fourier(phi(xx,tt)**3,xx),tt)$
listequa:={}$

%  The frequency w_1 is proportional to a(1)^2.

w1:=C_omega*a(1)**2$
for j:=2:n do  a(2*j-1):=c(2*j-1)*a(1)$
equation1:=equation1$

% Construct the system (7): R_{jj}(c)=0.

for k:=1:3*n do
    listequa:=append(listequa,
            {16*df(equation1,sin((2*k-1)*xx),
                             sin((2*k-1)*tt))/a(1)**3});

% Find a real solution of this system.

on rounded$
C_omega:=0$

% C_omega must be connected variable, because I want to clear it
% without message:"WARNING...".

for k:=2:n do
<< for j:=k:n do c(2*j-1):=0;
   clear C_omega, c(2*k-1);
   firstequa:=first(listequa);
   C_omega:=C_omega-firstequa/df(firstequa,C_omega);
   equa:=part(listequa,k);

% All equations, which we solve using the procedure SOLVE, are
% cubic equations, hence, they have at least one real root. We assume
% that the absolute value of one of real roots is less than unit.
% We solve the equation and select this real solution
% as value of c(2*k-1).

   solve_equa:=solve(equa,c(2*k-1));
   while solve_equa neq {} do
 << c(2*k-1):=part(solve_equa,1,2);
    if( c(2*k-1)=sub(i=-i,c(2*k-1)) and abs(c(2*k-1))<1)
      then solve_equa:={}
      else solve_equa:=rest(solve_equa);
 >>;
  test:=0;
  while(test=0) do
 << test:=1;
    clear C_omega;
    firstequa:=first(listequa);
    C_omega:=C_omega-firstequa/df(firstequa,C_omega);
    for j:=2:k do
  <<
    equa:=part(listequa,j);
    if(abs(equa)>10**(-11)) then
    << clear c(2*j-1);
       test:=0;
       equa:=part(listequa,j);
       solve_equa:=solve(equa,c(2*j-1));
       while solve_equa neq {} do
          << c(2*j-1):=part(solve_equa,1,2);
             if(c(2*j-1)=sub(i=-i,c(2*j-1)) and abs(c(2*j-1))<1)
               then solve_equa:={}
               else solve_equa:=rest(solve_equa);
          >>;
    >>;
  >>;
 >>;
>>;

% Write the obtained result.

c(1):=1$
for j:=1:n do  write "c(",2*j-1,"):=", c(2*j-1);
for j:=2:n do  write "c(",2*j-3,")/c(",2*j-1,"):= ", c(2*j-3)/c(2*j-1);
write "C_omega:=", C_omega;
for j:=1:3*n do  write "R(",j,",",j,"):=",part(listequa,j);
quit;

\end{verbatim}

The procedure "fourier" expands polynomials of $\sin(x)$ and
$\cos(x)$ in Fourier series.

\begin{verbatim}

  procedure fourier(FF,X);

% This procedure constructs Fourier-series expansions for polynomials
% of sin(x) and cos(x).

  begin
        scalar F;
        for all a, b such that df(a,x) neq 0 and df(b,x) neq 0
               let
                  cos(a)*cos(b)=(cos(a-b)+cos(a+b))/2,
                  sin(a)*sin(b)=(cos(a-b)-cos(a+b))/2,
                  sin(a)*cos(b)=(sin(a-b)+sin(a+b))/2,
                  sin(a)**2=(1-cos(2*a))/2,
                  cos(a)**2=(1+cos(2*a))/2;
        F:=FF;
        for all a, b such that df(a,x) neq 0 and df(b,x) neq 0
                clear
                     cos(a)*cos(b),
                     sin(a)*sin(b),
                     sin(a)*cos(b),
                     sin(a)**2,
                     cos(a)**2;
        return F;
  end;

\end{verbatim}

\end{document}